# Registration of the signal of a star and PCR sources optical radiation by means of the installation, aimed at the investigation of EAS of high energy cosmic rays

T.T.Barnaveli, N.A.Eristavi, I.V.Khaldeeva

Andronikashvili Institute of Physics, Tamarashvili 6, Tbilisi, 0177, Georgia

A.P.Chubenko, N.M.Nesterova

Lebedev Physical Institute, Leninsky av. 53, Moscow, Russia

T.T.Barnaveli (jr)

Inventors Network GmbH 22453, Kollau str. 123, Hamburg, Germany

E-mail: tengiz.barnaveli@gmail.com

## Abstract

With the help of the experimental installation aimed at the investigation of high energy cosmic rays (Tien-Shan high mountain laboratory) the signal of Solar and star optical radiation is registered. The signal is well provided statistically and possesses the strictly expressed maximum in the region of EAS sizes $N_e \approx 1.19 \cdot 10^6$ particles (primary energy $E_0 \approx 1.33 \cdot 10^{15}$ eV). This signal is the peak from gamma EAS, generated by gamma quanta from decay of $\pi_0$ mesons, photo produced by the Primary Cosmic Radiation (PCR) nuclei on the photons of stars and of PCR sources. The assumption is made, that exactly this process provides the main contribution in the formation of so called "knee" on the primary spectrum.

Due to the universality and distinct maximum of this signal, its usage for independent and reliable calibration of the EAS installations, for the mutual calibration of these installations and, possibly, for the merger of experimental data obtained by means of these installations to increase the statistics, is proposed. It is especially vital to day, when the further essential increase of the energy range under investigation is necessary.

***

The experimental material, obtained by means of the Tien-Shan high mountain installation in the 70's – 80's and treated anew taking into account the new tasks, was used. More than 300 000 events were included in the treatment.

Our goal is to find and register the signal of a well established object of interstellar medium, the object of the guaranteed properties. Such a signal, in particular, could serve as an important base for the independent and reliable calibration of the installations, aimed at investigation of high energy cosmic radiation. It is clear, that for these goals it would be better to have more than one separate reference mark. Now we are trying to find such a possibility - however at this stage we will try to provide at least one reliable and, what is important, universal reference point.

Most real and attractive seems to us to choose as such an object the optical radiation of the Sun, stars and PCR sources. This radiation possesses a rather sharp maximum in the energy region 3.2 **eV**. The searched signal will be the peak from gamma EAS generated by gamma quanta from the decay of $\pi_o$ mesons, photo produced by the nuclei of the primary cosmic radiation on the photons of the Sun, stars and PCR sources. As we will see hereafter, the EAS, generated by these gamma quanta, turn out to be located in the region $N_e \approx 1.19 \cdot 10^6$ particles (primary energy $E_o \approx 1.33 \cdot 10^{15}$ **eV**). This is namely the region where most of the EAS installations provide high sensitivity and good precision of data treatment. At the same time the photons created via $\pi_o$ decays possess much less energy than is necessary to maintain the development of an electromagnetic–nuclear cascade (this may take place just at the very first stages). This gives the possibility of very pure selection of these cascades with an account of requirement of high value of the age parameter S at the observation level, and due to much higher correspondence between their spatial distribution and NKG function. Starting namely with these considerations we have built our analysis.

The spectrum of solar radiation [1] is shown in Fig.1. As we have mentioned above, it possesses a sufficiently narrow peak in the region of 3.2 eV ($10^{14}$ Hz, wave length 500 nm). If one takes into account that our Sun is an ordinary "average" star, this spectrum may be taken as the basic for the analysis.

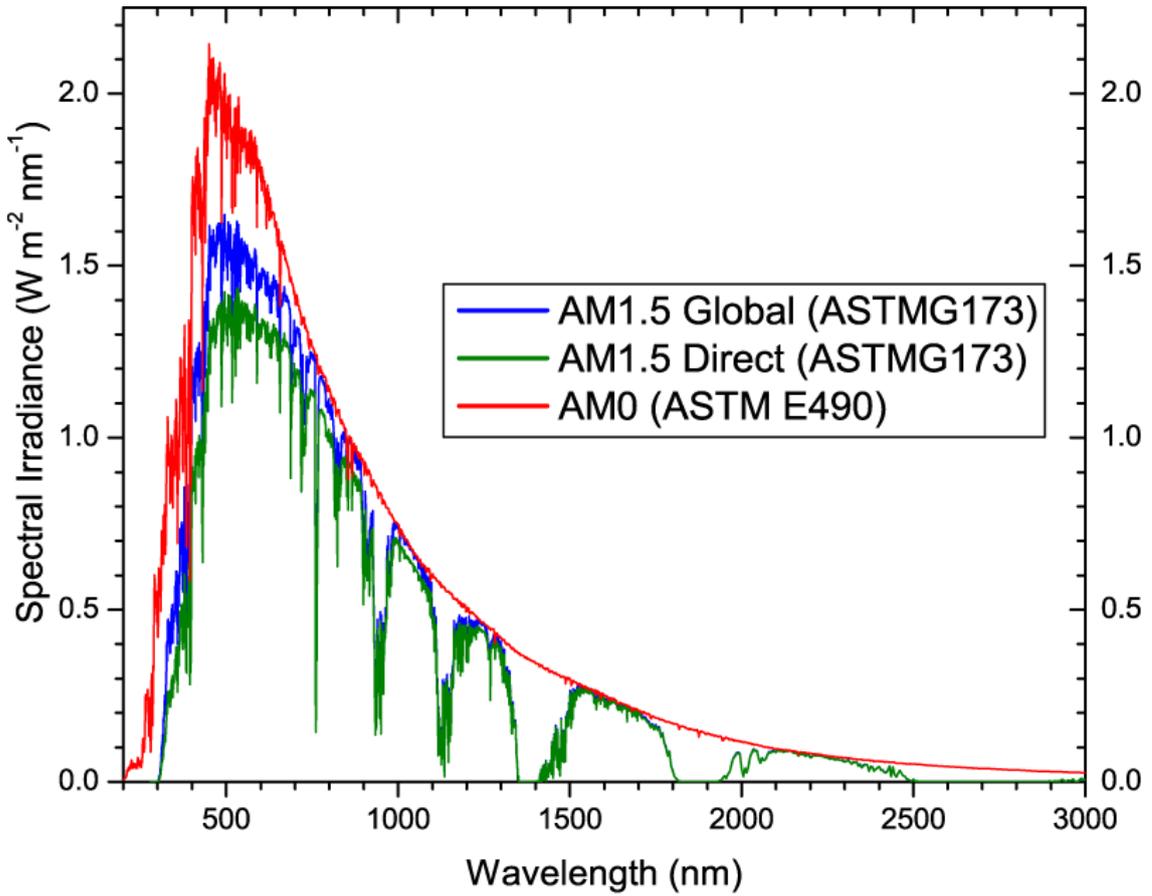

Fig.1. Standard Solar Spectra for space (AM0, the largest contour) and terrestrial (AM1.5) use.

The calculation of $\pi_0$ photo production was carried out in accordance with the work [2]. In the laboratory system the threshold energy $\omega_{th}$, necessary to create the particle, say, of the mass $m_\pi$ is :

$$\omega_{th} = ((m_f + m_\pi)^2 - m_i^2 + Q^2) / 2\, m_i \qquad (1)$$

where

$m_f$ is the mass of the nucleon target in final state,

$m_i$ is the mass of the nucleon target in initial state,

$Q^2$ is the transmitted four momentum square, which equals to zero in our case of the real photon.

This searched energy of photons turns out to be 144.68 MeV in the rest system of the proton – Table 1. This Table (more correctly its string, concerning $\omega_{th}$) is taken from the work [2] and is expanded according to the requirements of our task.

| Incident nucleon | π+ | π− | π₀ | η |
|---|---|---|---|---|
| | Threshold energies $\omega_{th}$ of these meson photo production (all energies in this Table are given in **eV**) | | | |
| P | 1.5143·10⁸ | − | 1.4468·10⁸ | 7.093·10⁸ |
| N | − | 1.4845·10⁸ | 1.4467·10⁸ | 7.091·10⁸ |
| γ– factor and energy of the incident nucleon | | | 0.226·10⁸  2.121·10¹⁶ | 0.858·10⁸  8.046·10¹⁶ |
| γ– factor and energy of the created meson | | | 1.976·10⁷  2.667·10¹⁵ | 5.410·10⁷  2.969·10¹⁶ |
| The energy of the separate gamma quantum, created in meson decay | | | 1.334·10¹⁵ | 1.485·10¹⁶ |
| The number of particles in EAS, generated by this gamma quantum | | | 1.19·10⁶ | 1.90·10⁷ |

In order to make the star light photon of the energy $\omega_o$ =3.2 **eV** possess (in the rest system of the incident proton) the energy ω=144.68 **MeV**, the proton due to the Doppler effect must move with the γ – factor $\gamma = \omega/2\omega_o = 0.226·10^8$.

Starting from the quoted evaluation one can find the energy of incident proton in the laboratory system $E_p = m_p · \gamma = 2.121·10^{16}$ eV.

So γ – factor of the whole born compound system will be [3]

$$\gamma = E_p / (m_p + m_\pi + \omega_{phot}) = 1.976·10^7$$

From here the energy $\omega_\pi$ of the created π₀ meson will be

$$\omega_\pi = \gamma · m_\pi = 2.667·10^{15} \text{ eV}.$$

The created $\pi_o$ mesons, whilst decaying give rise to the gamma quanta, which in turn create gamma showers registered by means of the installation. All numeric data are quoted in Table 1.

For the photo production of the next-by-mass meson $\eta$, $\gamma$ – factor of the incident proton must equal $\gamma = 0.8575 \cdot 10^8$, whilst its energy, correspondingly will be $E_p = m_p \cdot \gamma = 8.046 \cdot 10^{16}$ eV. This energy exceeds the energy, required for the creation of the $\pi_o$ meson approximately 4 times. Consequently the amount of such primary protons will be more than 60 times less, taking into account the primary cosmic radiation spectrum exponent $\xi = -2.7$. If additionally, one takes into account the partial share of decay in 2 $\gamma$ of the order of 40%, it means that the signal from $\eta$ meson will be too small (not to mention more heavy mesons).

One must stress, that $\pi_o$ mesons will be created by the nucleons of any nuclei moving with $\gamma$ – factor $\geq 0.226 \cdot 10^8$ (Table 1). That means that any nucleus, possessing the energy of the order of about of $2 \cdot 10^{16}$ eV /per nucleon, will contribute to this signal. The same for neutrons. This process is connected with the nucleons, but not with the nucleus, because it takes place at the energies much higher, than the binding energy of the nucleus.

Here most important is, that at any values of minimal and maximal energies the median energy of gamma quanta remains the same and equals:

- in the rest system of $\pi_o$ $\quad \omega_\gamma^* = p_\gamma^* = E_{\pi o} / 2 = m_{\pi o} / 2 = 67.5 \cdot 10^6$ eV,
- in the observation point $\omega_\gamma = m_{\pi o} / 2 \cdot \gamma_{\pi o} = 1.334 \cdot 10^{15}$ eV, i.e. is determined by $\gamma$ factor of the created $\pi_o$ meson. Note that this value lies approximately in the region of the well-known bump (or "knee") on the primary spectrum – we will return to this question further.

The experimentally registered signal, which we finally will take for the signal of the optical radiation interaction with the high energy primary cosmic radiation, will be extracted out of the general background namely starting from the assumption of its gamma nature and using all the accompanying effects.

The next considerations help us to extract this signal from the general spectrum of high energy EAS:

1. These gamma EAS must be close to purely electromagnetic EAS. Really, as we have noticed above, at these energies the gamma quanta participating in the development of cascade do not possess the energy high enough to generate and maintain the nuclear cascade via inelastic processes (may be just at the very first stages). So these EAS must be characterized by the age parameter **S** close to the maximal value **S**=2.
2. For the same reason these cascades must fit the NKG function well.

Taking for selection criteria the value of age parameter **S** in the frames 1.9 – 1.99, we obtained the signal shown in Fig. 2. This spectrum was built taking into account the acceptance of installation. Then, to flatten the graph, each point was multiplied by the corresponding value of $N_e^{**}2$. The result is shown in Fig. 3.

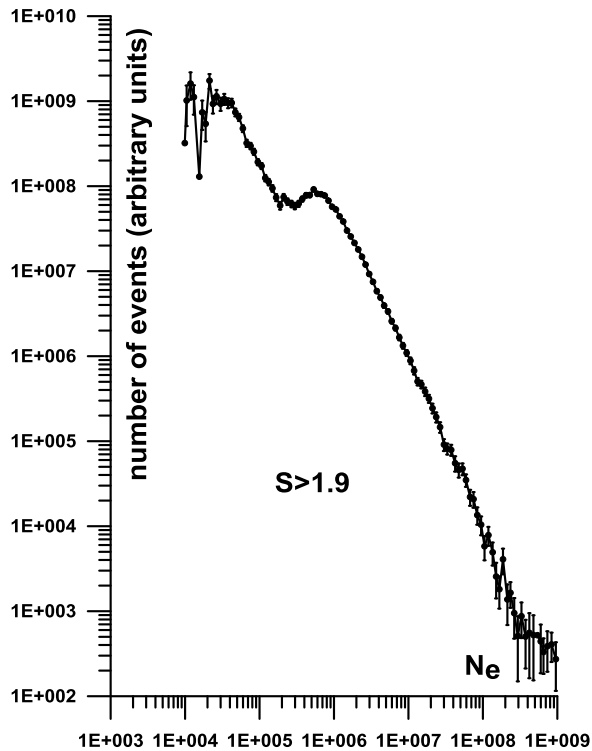
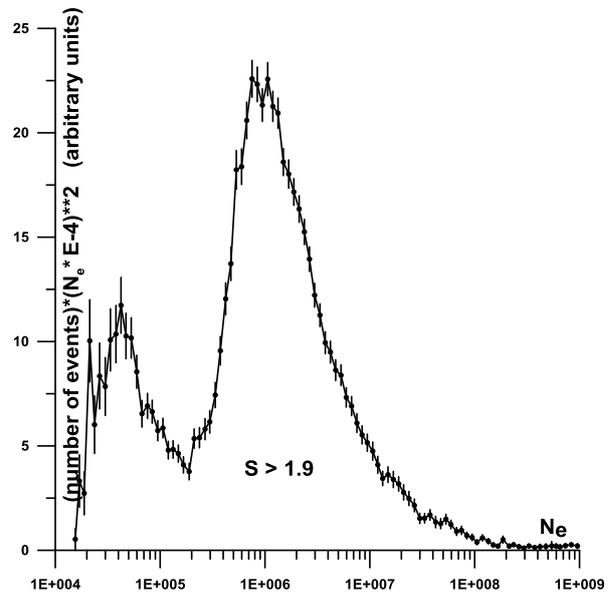

Fig.2                                            Fig.3

EAS spectrum with account the acceptance of installation (**S**>1.9)

The same spectrum, but each point was multiplied by the corresponding value of $N_e^{**}2$.

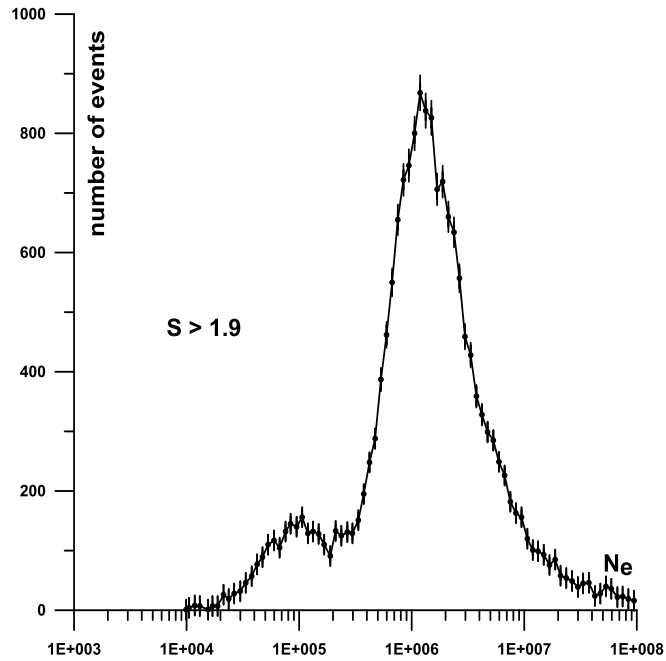

Fig. 4

The same spectrum, but built without any corrections

The same spectrum, but built without account of acceptance of installation and without multiplication by the **$N_e$**2, is shown in Fig. 4.

Of course, the more strict criteria may be applied as well (see below) - the quality of the experimental material and the level of data treatment achieved as at today, allow their application.

It is to be stressed that due to the high steepness of the primary spectrum, the main contribution in the formation of the signal under consideration comes from the region of the spectrum in the vicinity of the decaying particle ($\pi_o$ meson) creation threshold.

The transition from the EAS primary energy **$E_0$**, to the corresponding EAS size **$N_e$** and reverse, is calculated following the "Dunaevsky formula", that traditionally is used in Tien – Shan laboratory and was especially elaborated for this installation

$$\mathbf{N_e} = (\mathbf{E_0}\,\text{TeV}/0.015)^{**}(1/0.87) \qquad (2)$$

For the quoted value $E_0 = \omega_\gamma = 1.334 \cdot 10^{15}$ eV the size **$N_e$** of corresponding EAS turns out to be $4.9 \cdot 10^5$ particles. At the same time the experimentally obtained value is 2.44 times higher and is equal to $1.19 \cdot 10^6$ particles. Here the essence is that the expression (2) was derived long time ago for the "mean" EAS, including the showers over the whole spectrum of primary masses – starting with heavy nuclei and down to gamma quanta. Because of this, formula (2) lowers the value of $N_e$ at the observation level, due to the necessity to take into account the transit of the energy in showers - on account of the hadron and muon components, that is characteristic for the EAS caused by nucleons and nuclei. If one believes that the obtained signal is really what we think about it (i.e. the signal of gamma EAS of the primary energy $E_0 = \omega_\gamma = 1.334 \cdot 10^{15}$ eV) then the expression (2) is to be changed, to get the formula acceptable for gamma showers (in any case for the energy region under consideration). At this stage we just use the correction coefficient 2.44, in case of $\pi_o$ and $\eta$ mesons. Possibly the figure 0.87 in the exponent also may need to be corrected to make the formula applicable in the wide range of primary energies. In favor of the mentioned nature of the signal speaks the fact that **in the frames of the above mentioned restrictions of input parameters there are no other statistically significant signals in the energy borders in the Figures 2 - 9.**

One must note here, that at the further increase of the number and rigidity of restrictions (namely adding the mean square deviation **FF** of the spatial distribution function from NKG function and limitation of the energy exposure in the calorimeter i.e. the share of the hadron component), one can bring the experimental signal to complete correspondence with formula (2), and even shift it toward lower energies. The plateau at the left side of the main signal in Fig. 4 almost vanishes (it oscillates around the level of about 10 events). But this situation is not stable. The borders of this correspondence are very narrow. By their gradual widening, the maximum of signal rapidly and continuously shifts to the right, and at the condition mentioned above (**S** ≥ 1.9) we get the signal, shown in Fig. 4. Inside the borders of 1.8 < **S** <1.99 and **FF** ≤ 20% the position and the shape of the signal remains practically stable, and also stable is its amplitude with a precision of about 2%. This, from our point of view, speaks in favor of the signal's veritablity. And one more remark. Any restriction of the amount of hadrons in the calorimeter is not desirable - because it leads to selection of the qualitatively other showers; in particular the contribution of inelastic processes with small transverse moments is suppressed: the calorimeter controls mainly the central parts of the showers.

It is quite natural, that as a result of photo production, except for $\pi_0$ mesons, the charged $\pi$ mesons will be created practically of the same energies. In the Figure 4 the plateau to the left of the main signal evidently is connected exactly with the creation of charged $\pi$ mesons with their consequent decay into $\mu^{\pm}$ and then $e^{\pm}$. To this question we will return a little later.

As to the shape of the front and back sides of the signal, they are the rather complicated function of the $\gamma$ factors, nucleon-photon meeting angles and of the shape of the primary spectrum in each concrete point. However, for our goal it is not essential.

To make it visual, in the Figures 5 – 9 the EAS spectra are shown obtained under the application of sequentially more and more strict selection criteria (of the age parameter $S$), the last step of which was the spectrum shown on Figures 2, 3 and 4.

On Fig. 5 the EAS spectrum is presented, obtained without any limits on the value of parameter $S$, and on the fitting level to the NKG function. Only the distances of the EAS axis location from the center of the installation are limited (up to 150 m), and the zenith angles of EAS axis inclination are limited as well (up to 40 degrees).

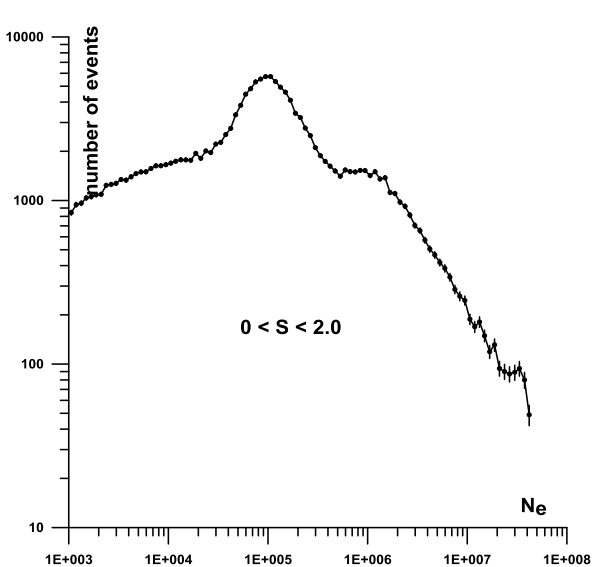
Fig. 5

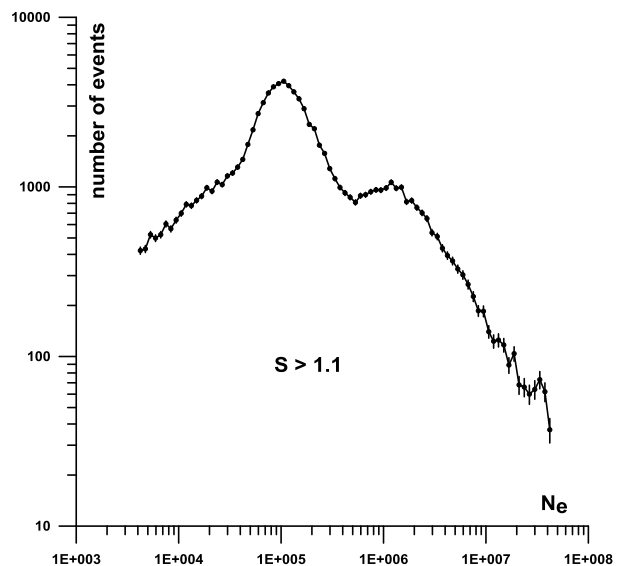
Fig. 6

In this picture, the distribution has the shape of a broad bell with the maximum in the region of EAS sizes of the order of $N_e = 10^5$ particles. The shape of distribution is defined by the joint influence of the falling primary spectrum of $N_e$ and the registration effectiveness increasing with $N_e$. This creates the presence of the maximum. However, it does not affect the local peculiarities. Exactly this spectrum was used to take into account installations geometry and acceptance, as was mentioned above. We introduced corresponding corrections starting with the generally accepted power form of the primary energy spectrum with the power index $\gamma = -2.7$.

The above mentioned "knee" may be clearly seen – the well-known peculiarity of the primary spectrum in the region $N_e$ of the order of $10^6$.

In the next figure 6, the restriction was applied: $S > 1.1$. Then, sequentially, the restriction lower limit of $S$ is increased by 0.2. In the Fig.7 ($S > 1.3$) the height of the peculiarity becomes close to the maximum, defined by the above mentioned influence of the installation acceptance. At $S > 1.5$ the peculiarity already preponderates (Fig. 8). At $S > 1.7$ (Fig. 9) just a hint of the left maximum remains.

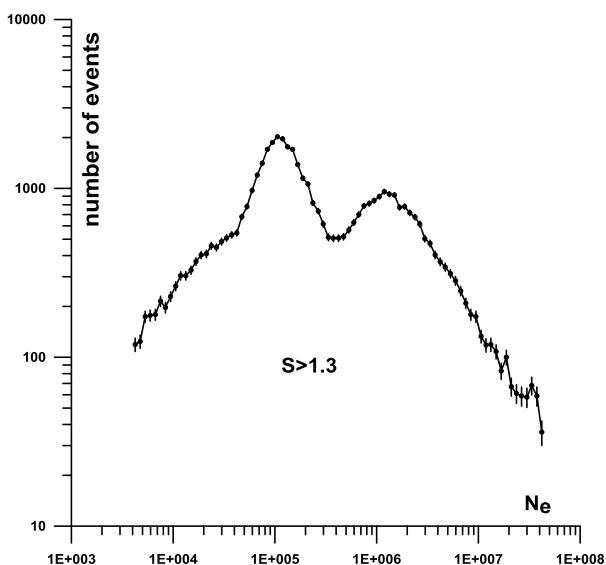 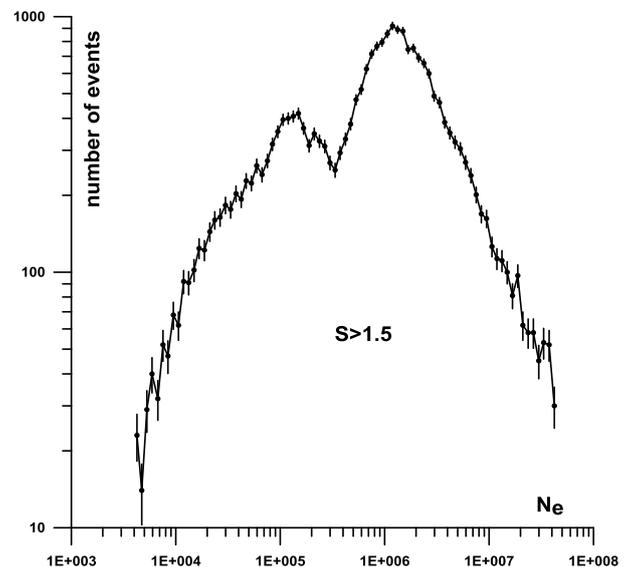

Fig. 7            Fig. 8

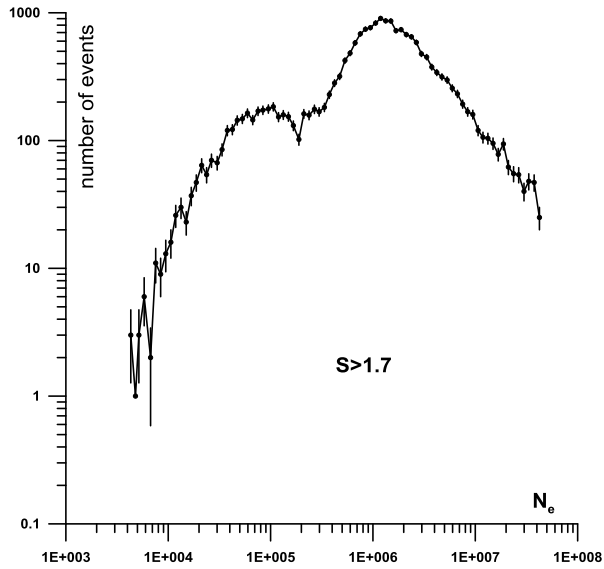

Fig. 9

Now it becomes convenient to use the linear scale for the vertical axis, and using the restriction **S** > 1.9 we get our main signal shown in Fig. 4. So the main selection criterion turns out to be the age parameter **S**, but one can use FF as well for additional reliability sake.

Note, that on all these figures the interval borders by $N_e$ are chosen in a way to make the main signal be located approximately in the center of the interval.

The share of the events contributing to the peak in Fig.4 (at its half height) in the corresponding energy region of Fig. 5 is about 40%.

**We have to stress once more that in the frames of the above mentioned restrictions of input parameters there are no other statistically significant signals in the energy borders in the Figures 2 - 9.**

## Discussion

Let us try to evaluate the share of purely electromagnetic events generated in the process of $\pi_0$ meson photo production, and their consequent decay into gamma quanta. At the first stage, it seems to be natural to suppose the Sun as the basic source of the photons. We accept the following input parameters: the cross section $\sigma_{\gamma p} \approx 10^{-28} sm^2$ ; the density of 3.2 eV energy photons near the Earth one can get from the known flux $\varepsilon$ of solar radiation on the Earth $\varepsilon = 0{,}137\ \text{W/cm}^2 = 10^{19}\ \text{erg/cm}^2$; then photon medium density near Earth $\rho = \varepsilon/\omega c \approx 10^8\ \text{cm}^{-3}$.

Let **R** be the distance between the Sun and the Earth. Now it is possible to evaluate the probability of interaction of the proton with the photon medium at the distance **L** from the Sun, $P = \sigma_{\gamma p} \cdot \rho \cdot R^2 \cdot \int_R^\infty \frac{dL}{L^2} = \sigma_{\gamma p} \cdot \rho \cdot R^2 / L$. We are doing approximate evaluation, so for simplification sake we restricted the integration area by **R** from below, so excluding the area within the sphere of radius **R**. It is quite admissible, if one takes into account that the experimental intensities of signal in the day and in the night are close to each other within one standard deviation. This means that if the signal is really generated on the solar photons, then the relative probability for the primary nucleon to pass in the close vicinity of the Sun is quite small. The calculation leads to the result of at least 3 orders lower than the experimentally observed intensity. Note, that in the calculation was additionally inserted the correction taking into account the primary cosmic ray mass composition in the energy region around the $\approx 10^{16}$ eV, where the mean mass of the primary nuclei is **M**=12. Then we tried to evaluate the possible contribution of the optical irradiation of all of the stars of the Galaxy. If one starts with the accepted life time of the proton in our Galaxy (the model with Galo) of the order of $10^8$ years (protons of the energy $\geq 10^{17}$ already are not captured by the Galactic magnetic fields) and with the mean distance between the stars in the Galaxy of the order of 6 light years, then the proton becomes able to "visit" the vicinity of about $10^8$ stars with a mean distance about 2 light years, and with the probability close to 1 to cross at least once the photosphere of one of them. Attributing to all of the stars the characteristics of the Sun we nevertheless obtained a result of about two orders lower, than the experimentally registered share of these events. So the joint effect of the Sun and of $10^8$ stars of the Galaxy may compose not more than 1 – 2 % of the experimental result. So the conclusion will be as follows – in interstellar space the probability of $\pi_0$ production is not high enough to obtain the observed effect. Nevertheless the experimental result exists and is very well provided statistically. We do not see any other possibility than to accept, that the process takes place mainly already in the cosmic ray sources themselves, where the density of the optical radiation and the length of the route of the protons and nuclei in the outer shells are evidently very high.

In this connection let us return again to the figure 4, where except for the main peak we see plateau-like spectrum of gamma EAS, immediately adjoining the basic signal. The only real possibility to explain the emergence of such plateau is the synchrotron radiation of electrons and positrons (and possibly of muons, though they are too massive to contribute essentially), created in the consequent decays of $\pi^{\pm}$ mesons, photo produced (as the $\pi_0$ mesons) in the interactions of high energy protons and nuclei with the optical radiation of source. Actually the spectrum in the figure 4 represents a united signal, where the peak corresponds to photo production of $\pi_0$ - mesons, while the plateau – of $\pi^{\pm}$ - mesons. At numeric calculations, one can with satisfactory precision take the momentum and the energy of the created $\pi^{\pm}$ -mesons as equal to the corresponding parameters of $\pi_0$ - meson (see Tab.1). By these assumptions let us evaluate the maximal energies of $e^{\pm}$ and $\mu^{\pm}$, created in the consequent decays of $\pi^{\pm}$ -mesons. Calculation was carried on in correspondence with formula

$$\omega_{min}^{max} = (\omega_0 \cdot \omega_1^* + p_0 \cdot p_1^*)/m_0 \qquad (3)$$

[3], where index 0 is applied for the decaying particles, index 1 – for the created particles, and sign * - for the rest system of the decaying particle. Calculations lead to the maximal energy of $e^{\pm}$ and $\mu^{\pm}$, equal to $\approx 2.64 \cdot 10^{15}$ eV. The corresponding number of the particles in EAS $N_e(E_0) \approx 2.6 \cdot 10^6$. So the right side of distribution of EAS from the synchrotron radiation immediately adjoins the basic signal and even intersects with its left front and overlaps with the signal, not exceeding its right side on the level of plateau (even in case of the simultaneous throw off of the whole energy, which hardly is probable).

Let us evaluate the magnetic field (its component, normal to the trajectory of the particle), able to provide the synchrotron radiation of the energy of the order of $2.67 \cdot 10^{15}$ eV. As it is known, the position of the maximum of the synchrotron radiation spectrum is connected with the magnetic field H by the expression

$$\nu \text{ (Hz)} = 4.6 \cdot 10^6 \text{ H } \omega^2 \text{ (eV)} \qquad (4)$$

For the gamma quantum of the energy $2.67 \cdot 10^{15}$ eV, the corresponding frequency is equal to $10^{28}$ Hz. So, according to (4), one can evaluate the acting magnetic field:

$$H = 1.5 \cdot 10^4 \text{ Oe}$$

Such fields (and higher ones) are characteristic (and in some cases are evaluated) for some star objects. So the appearance of such fields in the cosmic ray sources is quite natural.

## Conclusion

1. With the help of the experimental installation aimed at the investigation of high energy cosmic rays (Tien-Shan high mountain laboratory) the signal of star and PCR sources optical radiation is registered. The signal is well provided statistically and possesses the strictly expressed maximum in the region of EAS sizes $N_e \approx 1.19 \cdot 10^6$ particles (primary energy $E_0 \approx 1.33 \cdot 10^{15}$ eV).

2. The generation mechanism of this signal is the photo production of $\pi_0$ mesons by the nuclei of primary cosmic radiation on the photons of the star and PCR source optical radiation, followed by their decay on gamma quanta and generation of gamma EAS. This process is completely analogous to the **GZK** process, which takes place on the photons of the relic radiation.

3. We think that exactly this process provides the main contribution in the formation of the so called "knee" on the primary spectrum. Extracting the events with S > 1.9 one obtains the spectrum without the "knee" – Fig. 10. One can propose other processes as well, for example [4], however their contribution seems to be essentially lower (small "remnants" in Fig. 10 in the place of the "knee" – compare with the Fig. 5). This question deserves additional study.

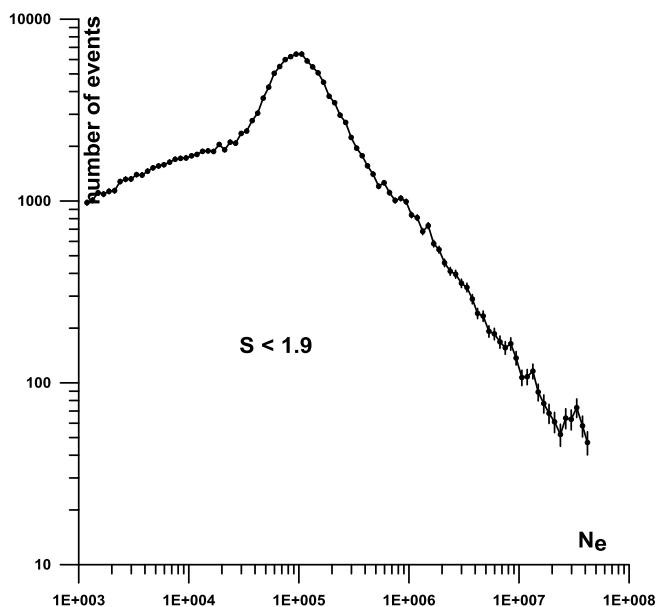

Fig. 10

The EAS spectrum after extraction of events with S > 1.9

4.  Due to the universality and sharpness of this signal its usage for independent and reliable calibration of the EAS installations, for the mutual calibration of these installations and, possibly, for the merge of experimental data obtained by means of these installations, to increase the total statistics is proposed. It is especially vital to day, when the further essential increase of the energy range under investigation is necessary.

## Acknowledgements.

The authors express their deep gratitude to O.V.Kancheli for his constant attention to our investigations and for fruitful discussions and remarks. The authors are sincerely grateful to J.M.Henderson for interesting discussions and his assistance in preparation of this paper.

## References.